\documentclass[apl,twocolumn]{revtex4}
\usepackage{graphicx}
\begin{document}

\title{Multistability and localization in coupled nonlinear split-ring resonators}

\author{Nikos Lazarides$^{1}$, Mario I. Molina$^{2}$, George P. Tsironis$^{1}$
        and Yuri S. Kivshar$^{3}$}

\affiliation{
$^{1}$Department of Physics, University of Crete and Institute
of Electronic Structure and Laser,
Foundation for Research and Technology-Hellas, P.O. Box 2208, 71003 Heraklion, Greece\\
$^{2}$Departamento de F\'isica, Facultad de Ciencias, Universidad
de Chile, Casilla 653, Santiago, Chile \\
$^{3}$Nonlinear Physics Center, Research School of Physics
and Engineering, Australian National University, Canberra ACT 0200,
Australia
}
\date{\today}

\begin{abstract}
We study the dynamics of a pair of nonlinear split-ring resonators (a `metadimer')
excited by an alternating magnetic field and coupled magnetically.
Linear metadimers of this kind have been
recently used as the elementary components for three-dimensional
metamaterials or 'stereometamaterials'
[N. Liu {\em et al}, Nature Photon. {\bf 3}, 157 (2009)].
We demonstrate that nonlinearity
offers more possibilities with respect to real-time tunability
and a multiplicity of states
which can be reached by varying the external field.
Moreover, we demonstrate almost total localization of the energy
in one of the resonators in a broad range of parameters.
\end{abstract}


\maketitle

Metamaterials are artificially structured composites which exhibit electromagnetic properties
not available in naturally occurring materials. Such structures are largely based on subwavelength 
resonant 'particles' refered to as
split-ring resonators (SRRs). Scaling down the size of the SRRs
allows to realize metamaterials up to Terahertz and optical
frequencies \cite{Linden-Soukoulis-Shalaev}.
However, for exploiting the great potential of metamaterials for applications,
it is desirable to change their effective parameters in real time, i.e.,
to achieve real-time tunability. This is also motivated the construction of nonlinear SRRs,
whose structure is very well suited for enhancing nonlinear phenomena
\cite{Pendry}. An effective way of constructing an easily controllable
nonlinear SRR is the insertion of a nonlinear
electronic component into its slit~\cite{Powell,Shadrivov2006,Wang}.
The arrangement of a large number of nonlinear SRRs into a periodic lattice
results in a nonlinear magnetic metamaterial which is tunable by varying
the power of the applied field \cite{Shadrivov2008}.

It was recently suggested that dimers, comprised of two
spatially separated SRRs,
can be used as elemental units for the construction of three
dimensional metamaterials ('stereometamaterials')~\cite{Liu2009}.
The close proximity of the SRRs in the
dimer results in relatively strong coupling between them.
A metamaterial comprised of a large number of
such metadimers can be utilized as a tunable optically active medium
\cite{Liu2007}.
Moreover, if one or both SRRs in the metadimer become nonlinear,
the metamaterial itself acquires nonlinear properties, essential
for real-time tunability.
It is therefore of great importance to investigate
the nonlinear properties of the elementary unit of such a
dimer-based metamaterial, i.e., of the nonlinear metadimer,
which can be modeled as system of two coupled nonlinear oscillators.

We consider an asymmetric metadimer comprised of two SRRs which have
slightly different slit widths as in Fig. 1, $d_{g,1}$ and $d_{g,2}$, which
differentiates their linear capacitances $C_1$ and $C_2$, respectively
\cite{Gorkunov,Molina}. That in turn differentiates the linear resonance
frequencies of the SRRs through the relation $\omega_i \simeq 1/\sqrt{L\, C_i}$
($i=1,2$), while the changes in the inductance $L$ and the Ohmic resistance $R$
of the SRRs are of higher order.
The relative orientation of the SRRs in the metadimer
can be such that the SRRs are either narrow-side coupled or broad-side coupled
(right and left panel of Fig. 1, respectively).
The nature of the interaction between the SRRs (electric or magnetic or both)
is determined by the twist angle of the slit of one of the SRRs around the
$\bf H$ field with respect to the other.
In both geometries shown in Fig. 1, that angle is $180^o$ so that the distance
between the slits is much larger than their widths.
Then, the interaction between
the SRRs is predominantly magnetic and the electric dipole-dipole interactions
can be neglected \cite{Hesmer,Penciu}.

\begin{figure}[h!]
\includegraphics[scale=0.3,angle=0]{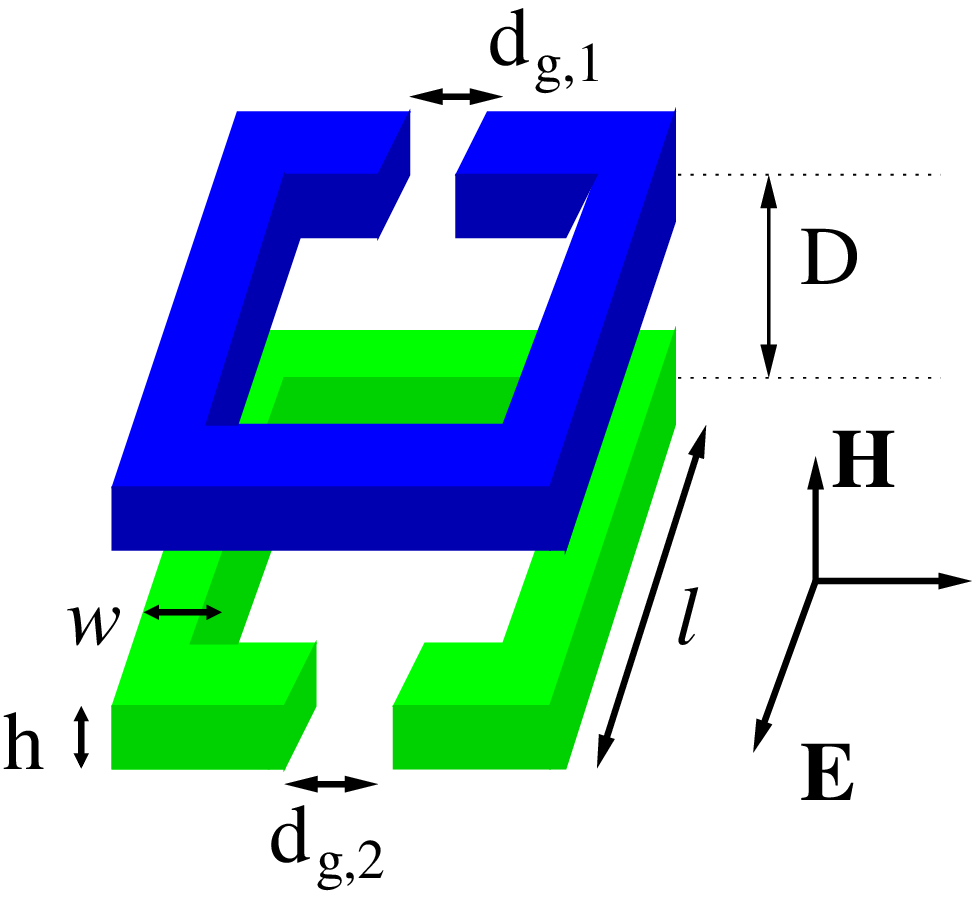}
\includegraphics[scale=0.3,angle=0]{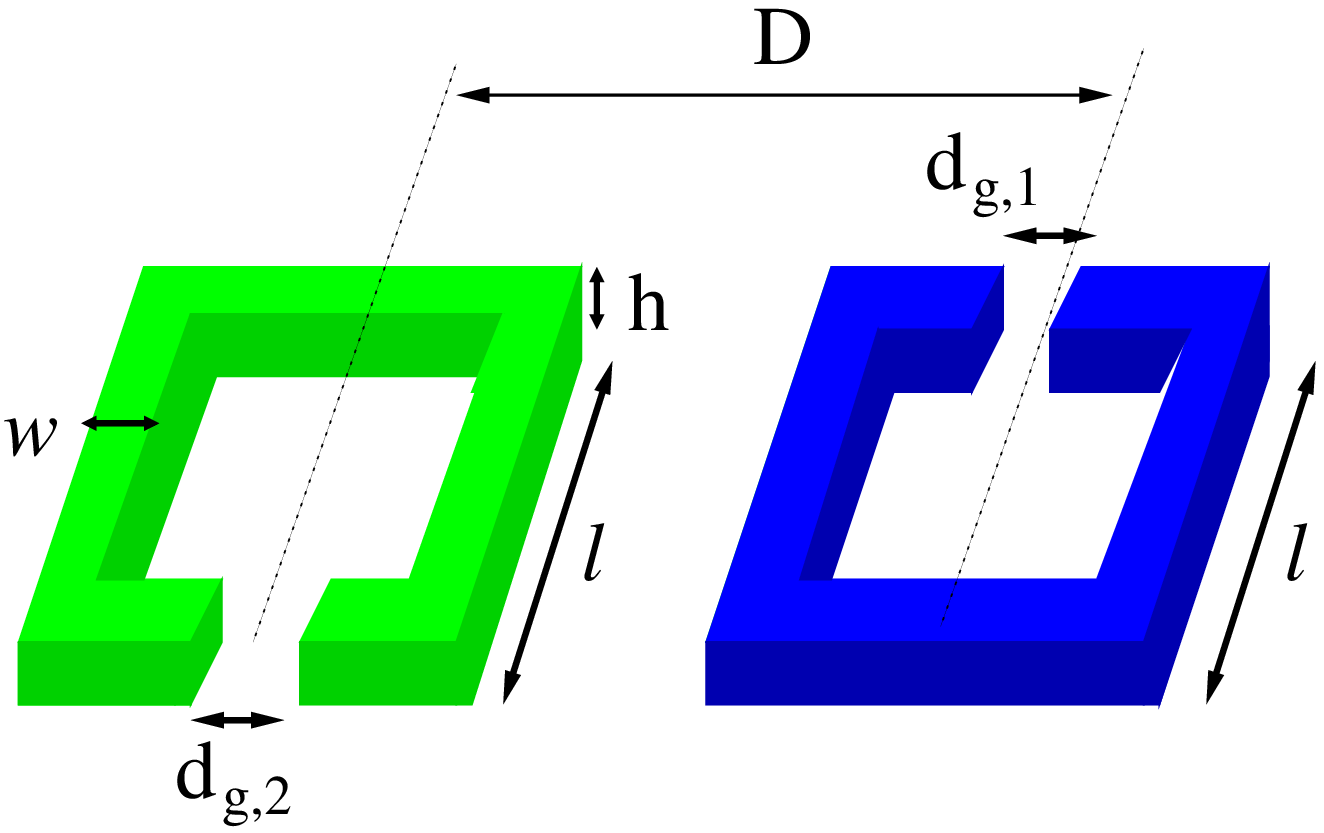}
\caption{
(color online)
Schematic of a broad-side (left panel) and narrow-side (right panel)
asymmetric metadimer.
}
\end{figure}

The metadimer is placed in an alternating electromagnetic field with the
polarization shown in Fig. 1, so that only its magnetic component is
capable of exciting induced currents in the SRR rings.
In the equivalent circuit picture the metadimer is regarded as a pair of
periodically driven, nonlinear resistor-inductor-capacitor
(RLC) oscillators coupled magnetically through their mutual inductance $M$,
driven by identical voltage sources.
Consider a metadimer with the geometry shown in the left panel of Fig. 1,
for which the magnetic coupling between the SRRs is relatively strong.
Then, the equations describing the dynamics of the (normalized) charge $q_i$,
accumulated across the slit of the $i-$th oscillator, reads \cite{Molina}
\begin{eqnarray}
 \label{1}
   \ddot{q_1} +\lambda_M \ddot{q_2} +\gamma \dot{q_1}
   + ( \partial u_1 /\partial q_1 )
   = \varepsilon_0 \sin(\Omega \tau) \\
 \label{2}
   \ddot{q_2} +\lambda_M \ddot{q_1} +\gamma \dot{q_2}
   + ( \partial u_2 /\partial q_2 )
   = \varepsilon_0 \sin(\Omega \tau) ,
\end{eqnarray}
where $\lambda_M =M/L$ is the magnetic interaction strength,
$\omega_0=\sqrt{\omega_1 \omega_2}$ a characteristic frequency,
$\varepsilon= \varepsilon_0 \, \sin(\Omega t)$ is the induced electromotive (emf)
force excited in each SRR,
$\gamma=R \sqrt{ {\sqrt{C_1 C_2}}/{L} }$ is the damping constant,
the overdots denote derivation with respect to the normalized temporal
variable $\tau$, and
\begin{eqnarray}
 \label{3}
  u_1=\frac{\delta}{2} q_1^2 \left( 1 -\frac{\delta^2}{2} \chi q_1^2 \right)
  \qquad
  u_2=\frac{1}{2\delta} q_2^2 \left( 1 -\frac{1}{2\delta^2} \chi q_2^2 \right) ,
\end{eqnarray}
with $\delta \equiv \omega_1 / \omega_2$ being the resonance frequency mismatch
(RFM) parameter, which quantifies the asymmetry of the metadimer.
The average energy, $E_{tot}$, of the metadimer can be obtained from the
time-average in one period of the Hamiltonian
\begin{eqnarray}
\label{10}
  H = \frac{1}{2(1 -\lambda_M^2)} \left( p_1^2 + p_2^2 -2\lambda_M p_1 p_2 \right)
  +u_1 + u_2 ,
\end{eqnarray}
where $p_1 = \dot{q}_1 + \lambda_M \dot{q}_2$ and
$p_2 = \dot{q}_2 + \lambda_M \dot{q}_1$,
with $p_i$ and $q_i$ calculated from Eqs. (\ref{1}) and (\ref{2}).
For normalizing the earlier equations, we have scaled
charge, voltage, time, and frequency by $Q_c$, $U_c$, $\omega_0^{-1}$,
and $\omega_0$, respectively, where
$Q_c=\sqrt{C_1 C_2} U_c$, $U_c=\sqrt{d_{g,1} d_{g,2}} E_c$,
with $E_c$ a characteristic electric field amplitude.

The rich dynamical behavior of the nonlinear metadimer can be observed
in Fig. 2 where a typical bifurcation diagram of the currents $i_1$ and $i_2$
is shown as a function of the driving frequency $\Omega$.
That diagram can be divided into three regions; the region at left,
where two stable periodic solutions coexist for a wide $\Omega$ interval,
the region at right where there is a single stable solution
(except in a narrow $\Omega$ interval), and the chaotic region
in between separating the previous two ones.
Multistability exists from $\Omega=0.6$
to $0.87$, where there are two stable states with very different energies;
the high and low energy state with $E_{tot} = 1.43$ and $E_{tot} =0.014$,
respectively.
Importantly, those two
states differ considerably in the distribution of $E_{tot}$ in the
two oscillators. We define the energy fractions in oscillators $1$ and $2$
as $e_1 = E_1/E_{tot}$ and $e_2 = E_2/E_{tot}$, respectively,
where $E_1$ and $E_2$ are their energies. The high
energy state has $e_1=0.993$ and $e_2=0.007$ while the low energy state
has $e_1=0.57$ and $e_2=0.43$. Thus, in the former case,
almost all the energy is localized in the first of the oscillators.

\begin{figure}[t!]
\includegraphics[scale=0.4,angle=0]{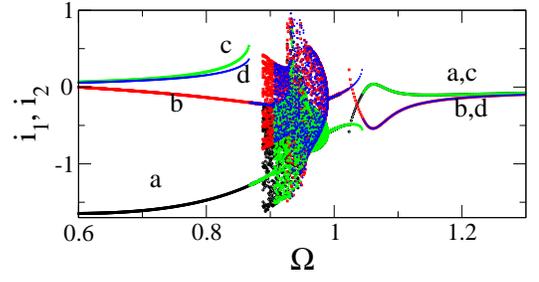}
\caption{
(color online)
Bifurcation diagram of the currents $i_1$ and $i_2$
as a function of the driving frequency $\Omega$,
for $\delta=0.95$, $\gamma=0.01$, $\chi=+1/6$, $\lambda=0.12$,
and $\varepsilon_0 =0.06$.
Decreasing $\Omega$: $i_1$ and $i_2$ correspond to black and
red symbols ($a$ and $b$), respectively; 
increasing $\Omega$: $i_1$ and $i_2$
correspond to green and blue symbols ($c$ and $d$), respectively.
}
\end{figure}

\begin{figure}[h!]
\includegraphics[scale=0.4,angle=0]{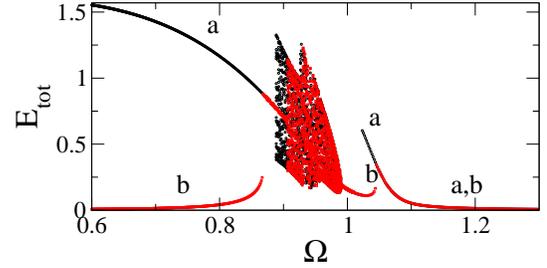}
\caption{
(color online)
Total energy of the metadimer $E_{tot}$
as a function of the driving frequency $\Omega$,
for the parameters in Fig. 2, and
$\Omega$ decreasing from $1.3$ (black symbols - $a$ );
$\Omega$ increasing from $0.6$ (red symbols - $b$).
}
\end{figure}

The total energy $E_{tot}$ of the two states and the energy fractions
$e_1$ and $e_2$ of the two oscillators as a function of $\Omega$
are shown in Figs. 3 and 4, respectively.
In that specific case, extreme localization occurs in a rather wide $\Omega$
interval, at least from $0.6$ to $0.87$.
If the metadimer is initially driven with low frequency it settles into the
low energy state. As the frequency increases, it passes through the point
where that state becomes unstable (at $\Omega=0.865$), and the metadimer
suddenly switches to the high energy state.
We also note that for $\Omega=0.885$ to $0.895$ a stable
periodic state coexist with a chaotic state (blue in red and green in black).

In the right-most region of Fig. 2, there is also a narrow $\Omega$
interval (from $\Omega=1.02$ to $1.05$) where multistability occurs.
Outside that interval, i.e., from $\Omega=1.05$ to $1.3$ the dimer state
has very low energy (see Fig. 3).
However, there are two specific values of $\Omega$,
i.e., at $\Omega=1.08$ and $1.025$, where $E_{tot}$ is highly localized.
At those points the energy fractions are $e_1=0.02$, $e_2=0.98$
and $e_1=0.99$, $e_2=0.01$, respectively.
Notice that the state at $\Omega=1.025$ lies in the region of bistability
in this part of the diagram, which has the lowest energy, while it also
exhibits localization. The corresponding higher energy state
at this frequency has $e_1=0.46$ and $e_2=0.54$,
so that $E_{tot}$ is almost
equally distributed in the two oscillators (see also Fig. 4).

\begin{figure}[h!]
\includegraphics[scale=0.4,angle=0]{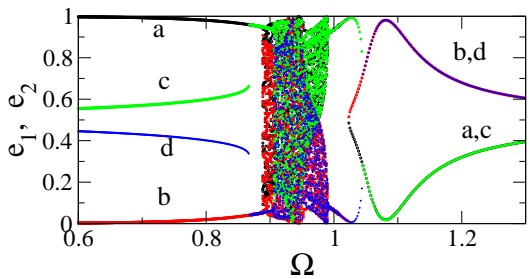}
\caption{
(color online)
Energy fractions $e_1$ and $e_2$ as a function
of the driving frequency $\Omega$ for the parameters in Fig. 2.
Decreasing $\Omega$: $e_1$ and $e_2$ correspond to black and
red symbols ($a$ and $b$), respectively; 
increasing $\Omega$: $e_1$ and $e_2$
correspond to green and blue symbols ($c$ and $d$), respectively.
}
\end{figure}

The currents $i_1$ and $i_2$ for $\Omega=0.7$ ($T=8.98$), which lies in
the region where both multistability and localization occur,
are shown in Fig. 5 as a function of $\tau$, both for the high (Fig. 5a) and the
low (Fig. 5b) energy states.
In the same figures we also plot $\cos(\Omega \tau)$
which is directly proportional to the applied magnetic field.
The relative phase difference between the applied
magnetic field and the current in each SRR, which is directly proportional
to its magnetic moment, determines the response of the metadimer to that field.
We observe that the currents of the metadimer in the high energy state
have a relative phase difference of almost $180^o$
with respect to the applied magnetic field, indicating a {\em diamagnetic}
magnetic response.
\begin{figure}[t!]
\includegraphics[scale=0.42,angle=0]{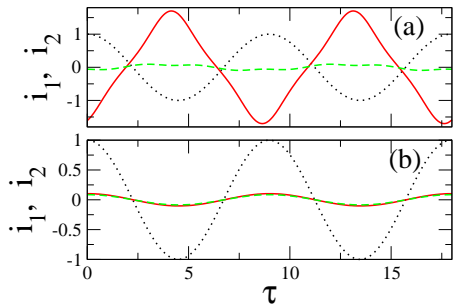}
\caption{
(color online)
Temporal evolution of the currents at $\Omega=0.7$,
for (a) the high energy state;
(b) the low energy state.
The other parameters are as in Fig. 2.
}
\end{figure}
To the contrary, the currents of the metadimer in the low energy state
are in phase with the applied magnetic field, indicating a {\em paramagnetic}
response.

In conclusion, we have revealed a physical mechanism of the intrinsic localization of energy 
in nonlinear magnetic metamaterials in the study of the dynamics of coupled sprit-ring resonators.
In particular, we have found that extreme localization of energy may occur in a slightly asymmetric 
nonlinear metadimer, which also exhibits multistability of states and chaos
for a rather wide range of parameters.  For a symmetric metadimer (with $\delta=1$ and other parameters 
as in Fig. 2), no localized states and chaos have been observed, while multistability still occurs.
Thus, it seems that a slight asymmetry is required for localization to occur.

We have found that, in the multistability regions, there appear two stable states which differ
considerably in their energies. The magnetic response of the metadimer
is either paramagnetic or diamagnetic, depending on the energy of the state
of the metadimer (low and high, respectively).
The magnetic response of a magnetic metamaterial comprised of such metadimers
is determined by averaging the response of the individual metadimers.
When the metadimers are in a high energy state, they may respond extremely
diamagnetically to an applied field. Then, the response of the magnetic metamaterial would 
be described macroscopically by a relatively large and negative magnetic permeability parameter.

We believe that this study may be useful for realizing strongly nonlinear 
effects and energy localization
in metamaterials comprised of a large
number of nonlinear split-ring resonators where strongly localized states 
in the form of discrete breathers
\cite{Lazarides2006,Eleftheriou2008,Lazarides2008,Eleftheriou2009},
as well as other types of localized excitations such as domain walls and
envelope solitons \cite{Shadrivov2006b,Kourakis} were predicted theoretically.

Y.K. acknowledges a support of the Australian Research Council 
and useful discussions with H. Giessen. 
M.I.M. acknowledges support from Fondecyt Grant 1080374.

\end{document}